\documentclass[twocolumn]{emulateapj}
\usepackage{epstopdf}

\newcommand{\myemail}{Wystan.Benbow@mpi-hd.mpg.de}
\newcommand{\hisemail}{Berrie.Giebels@poly.in2p3.fr}
\newcommand{\MPIK}{Max-Planck-Institut f\"ur Kernphysik, Heidelberg, Germany}
\newcommand{\Yerevan}{Yerevan Physics Institute, Armenia}
\newcommand{\Tolouse}{Centre d'Etude Spatiale des Rayonnements, CNRS/UPS, Toulouse, France}
\newcommand{\Hamburg}{Universit\"at Hamburg, Institut f\"ur Experimentalphysik, Germany}
\newcommand{\Berlin}{Institut f\"ur Physik, Humboldt-Universit\"at zu Berlin, Germany}
\newcommand{\Meudon}{LUTH, UMR 8102 du CNRS, Observatoire de Paris, Section de Meudon, France}
\newcommand{\Durham}{University of Durham, Department of Physics, U.K.}
\newcommand{\Potch}{Unit for Space Physics, North-West University, Potchefstroom, South Africa}
\newcommand{\LLR}{Laboratoire Leprince-Ringuet, IN2P3/CNRS,
Ecole Polytechnique, Palaiseau, France}
\newcommand{\Annecy}{Laboratoire d'Annecy-le-Vieux de Physique des Particules, IN2P3/CNRS, Annecy-le-Vieux, France}
\newcommand{\APC}{APC, Paris, France}
\newcommand{\DIAS}{Dublin Institute for Advanced Studies, Ireland}
\newcommand{\LSW}{Landessternwarte, Universit\"at Heidelberg, K\"onigstuhl, Germany}
\newcommand{\Montpellier}{Laboratoire de Physique Th\'eorique et Astroparticules, IN2P3/CNRS, Universit\'e Montpellier II, Montpellier, France}
\newcommand{\Saclay}{DAPNIA/DSM/CEA, CE Saclay, Gif-sur-Yvette, France}
\newcommand{\Grenoble}{
Laboratoire d'Astrophysique de Grenoble, INSU/CNRS, Universit\'e Joseph Fourier, Grenoble, France}
\newcommand{\Tuebingen}{Institut f\"ur Astronomie und Astrophysik, Universit\"at T\"ubingen, Germany}
\newcommand{\Jussieu}{LPNHE, IN2P3/CNRS, Universit\'es Paris VI \& VII, France}
\newcommand{\Prague}{Institute of Particle and Nuclear Physics, Charles University, Prague, Czech Republic}
\newcommand{\Bochum}{Institut f\"ur Theoretische Physik, Lehrstuhl IV: Weltraum und Astrophysik, Ruhr-Universit\"at Bochum, Germany
}\newcommand{\Namibia}{University of Namibia, Windhoek, Namibia}
\newcommand{\Erlangen}{Universit\"at Erlangen-N\"urnberg, Physikalisches Institut, Germany}
\newcommand{\LEA}{European Associated Laboratory for Gamma-Ray Astronomy, jointly supported by CNRS and MPG}
\newcommand{\PolandA}{Obserwatorium Astronomiczne, Uniwersytet Jagiello\'nski, Krak\'ow, Poland}
\newcommand{\PolandB}{Nicolaus Copernicus Astronomical Center, Warsaw, Poland}
\newcommand{\pks}{PKS\,2155$-$304}

\shorttitle{An Exceptional VHE Gamma-Ray Flare of PKS\,2155$-$304}
\shortauthors{Aharonian et al.}

\begin{document}
\title{An Exceptional VHE Gamma-Ray Flare of PKS\,2155$-$304}


\author{F. Aharonian\altaffilmark{1,2,14},
A.G.~Akhperjanian\altaffilmark{3},
A.R.~Bazer-Bachi\altaffilmark{4},
B.~Behera\altaffilmark{15},
M.~Beilicke\altaffilmark{5},
W.~Benbow\altaffilmark{2},
D.~Berge\altaffilmark{2,a},
K.~Bernl\"ohr\altaffilmark{2,6},
C.~Boisson\altaffilmark{7},
O.~Bolz\altaffilmark{2}
% ...
}

\altaffiltext{1}{Correspondence and request for material should be
  addressed to W. Benbow, \myemail, B. Giebels, \hisemail}
\altaffiltext{2}{\MPIK}
\altaffiltext{3}{\Yerevan}
\altaffiltext{4}{\Tolouse}
\altaffiltext{5}{\Hamburg}
\altaffiltext{6}{\Berlin}
\altaffiltext{7}{\Meudon}
% ...
\altaffiltext{14}{\DIAS}
\altaffiltext{15}{\LSW}
% ...
\altaffiltext{a}{now at CERN, Geneva, Switzerland}
% ...

\begin{abstract}
The high-frequency peaked BL\,Lac PKS\,2155$-$304 at redshift
$z=0.116$ is a well-known VHE ($>$100 GeV) $\gamma$-ray emitter.
Since 2002 its VHE flux has been monitored using the
H.E.S.S. stereoscopic array of imaging atmospheric-Cherenkov
telescopes in Namibia.  During the July 2006 dark period, the average
VHE flux was measured to be more than ten times typical values
observed from the object.  This article focuses solely on an extreme
$\gamma$-ray outburst detected in the early hours of July 28, 2006
(MJD 53944).  The average flux observed during this outburst is
I($>$200 GeV) = (1.72$\pm$$0.05_{\rm stat}$$\pm$$0.34_{\rm syst}$)
$\times$ 10$^{-9}$ cm$^{-2}$ s$^{-1}$, corresponding to $\sim$7 times
the flux, I($>$200 GeV), observed from the Crab Nebula.  Peak fluxes
are measured with one-minute time scale resolution at more than twice
this average value. Variability is seen up to $\sim$600 s in the
Fourier power spectrum, and well-resolved bursts varying on time
scales of $\sim$200 seconds are observed. There are no strong
indications for spectral variability within the data.  Assuming the
emission region has a size comparable to the Schwarzschild radius of a $\sim$$
10^9\,M_\odot$ black hole, Doppler factors greater than 100 are
required to accommodate the observed variability time scales.

\end{abstract}

\section{Introduction}

Flux variability studies provide a strong probe into the physical
processes of the innermost regions of Active Galactic Nuclei (AGN).
Although the broad-band emission from all AGN is highly variable, the
most extreme flux variability, i.e. largest magnitude and shortest
time scale, is observed from a class of AGN known as blazars. As a
result blazar variability studies are crucial to unraveling the
mysteries of AGN. Over a dozen blazars have been detected so far at
VHE energies.  In the Southern Hemisphere, PKS\,2155$-$304 is
generally the brightest blazar at these energies, and is probably the
best-studied at all wavelengths.  The VHE flux observed
\cite{HESS_2155A} from PKS\,2155$-$304 is typically of the order
$\sim$15\% of the Crab Nebula flux above 200 GeV.  The highest flux
previously measured in one night is approximately four times this
value and clear VHE-flux variability has been observed on daily time
scales. The most rapid flux variability measured for this source is
25\,min~\cite{HESS_2155B}, occurring at X-ray energies.  The fastest
variation published from any blazar, at any wavelength, is an event lasting
$\sim$800\,s where the X-ray flux from Mkn\,501 varied by 30\%
\cite[Xue \& Cui][]{MKN501_dispute}\footnote{Xue \& Cui~\cite{MKN501_dispute}
also demonstrate that a 60\% X-ray flux increase in $\sim$200\,s observed
\cite{MKN501_flare} from Mkn\,501 is likely an artifact.}, 
while at VHE energies doubling timescales as fast as
$\sim$15 minutes have been observed from Mkn 421 \cite{Gaidos_Mkn421}.

The High Energy Stereoscopic System \cite[H.E.S.S.;][]{HESS1} 
is used to study VHE $\gamma$-ray emission 
from wide variety of astrophysical objects.
As part of the normal H.E.S.S. observation program
the flux from known VHE AGN is monitored regularly
to search for bright flares.  During such flares, 
the unprecedented sensitivity of H.E.S.S.
(5 standard deviation, $\sigma$, 
detection in $\sim$30\,s for a Crab Nebula flux source at
20$^{\circ}$ zenith angle) enables studies of VHE-flux variability on
time scales of a few tens of seconds. During the July 2006 dark period,
the average VHE flux observed by H.E.S.S. from PKS\,2155$-$304 was
more than ten times its typical value.  In particular,
an extremely bright flare of PKS\,2155$-$304 was 
observed in the early hours of July 28, 2006 (MJD 53944).
This article focuses solely on this particular flare.
The results from other H.E.S.S. observations of 
PKS\,2155$-$304 from 2004 through 2006 will be published elsewhere.

\section{Results from MJD 53944}
\label{sect:results}

A total of three observation runs ($\sim$28 min each) were
taken on PKS\,2155$-$304 in the early hours\footnote{The three runs began at
00:35, 01:06 and 01:36 UTC, respectively.} of MJD 53944.  These data
entirely pass the standard H.E.S.S. data-quality selection criteria,
yielding an exposure of 1.32\,h live time at a mean zenith angle of
13$^{\circ}$.  The standard H.E.S.S. calibration \cite{calib_paper}
and analysis tools \cite{std_analysis} are used to extract the results
shown here.  As the observed signal is exceptionally strong, the
event-selection criteria \cite{std_analysis} are performed using the
{\it loose cuts}, instead of the {\it standard cuts}, yielding an
average post-analysis energy threshold of 170 GeV.  The {\it loose cuts}
are selected since they have a lower energy threshold and
higher $\gamma$-ray and background acceptance.  The
higher acceptances avoid low-statistics issues 
with estimating the background and significance
on short time scales, thus simplifying the analysis.
The on-source data are taken from a circular region of radius
$\theta_{cut}=0.2^{\circ}$ centered on PKS\,2155$-$304, and the
background (off-source data) is estimated using
the {\it Reflected-Region} method \cite{bgmodel_paper}.

A total of 12480 on-source events and 3296 off-source events
are measured with an on-off normalization of 0.215.  
The observed excess is 11771 events ($\sim$2.5Hz), 
corresponding to a significance of 168$\sigma$ calculated  following the 
method of Equation (17) in Li \& Ma~\cite{lima}.  It should be noted that 
use of the {\it standard cuts} also yields a strong excess 
(6040 events, 159$\sigma$) and results (i.e. flux, spectrum,
variability) consistent with those detailed later.

\subsection{Flux Variability}
\begin{figure}
\plotone{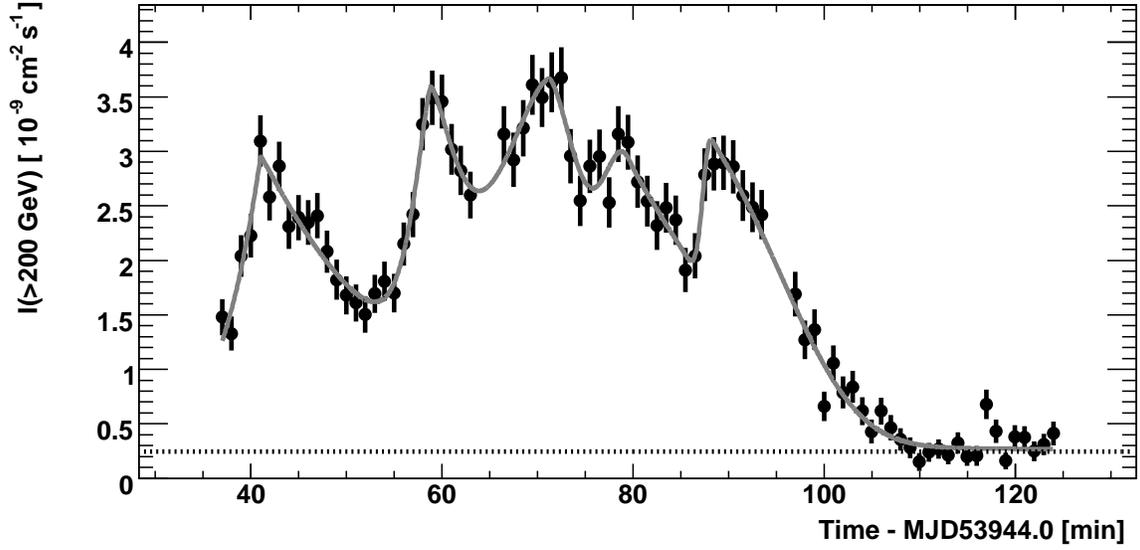}
\caption{The integral flux above 200 GeV observed
from PKS\,2155$-$304 on MJD 53944 versus time.  The data are binned
in 1-minute intervals.  The horizontal line represents
I($>$200 GeV) observed \cite{hess_crab}
from the Crab Nebula.  The curve is the fit
to these data of the superposition of five bursts (see text) and a
constant flux.\label{flux_lc_1min}}
\end{figure}

\begin{figure}
\plotone{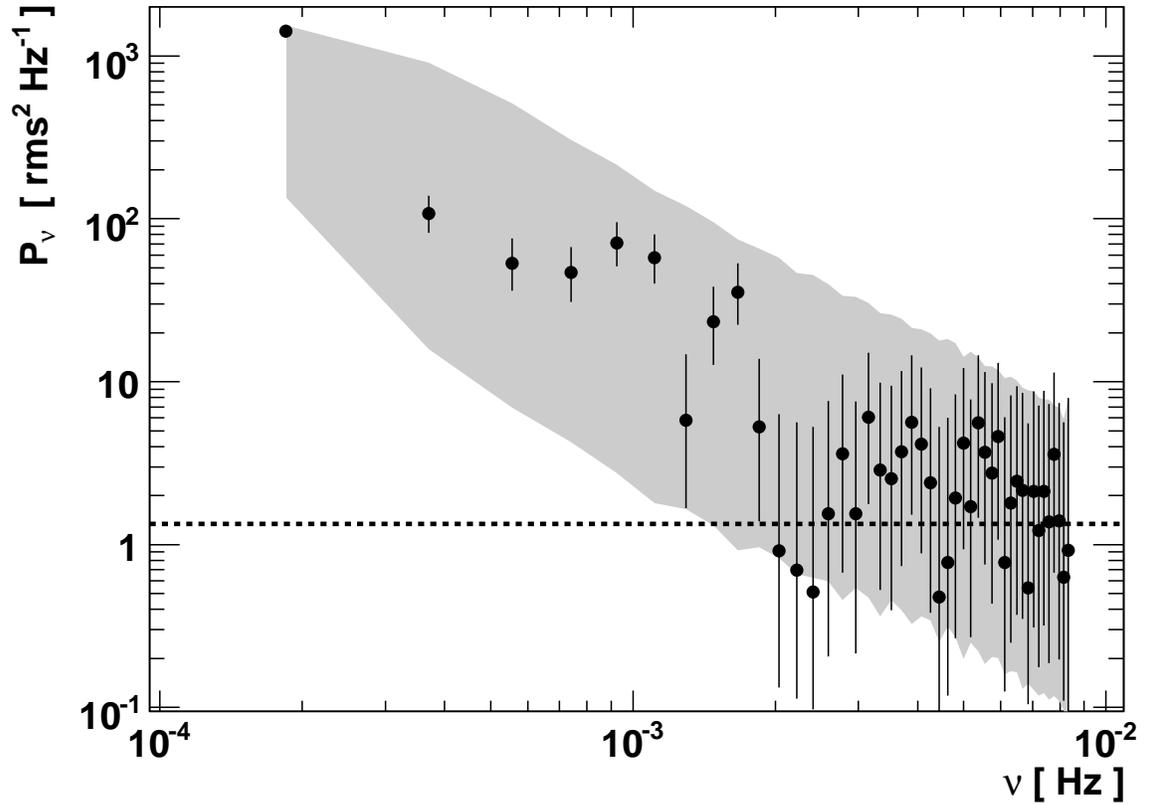}
\caption{The Fourier power spectrum of the
light curve and associated measurement error. The grey shaded area
corresponds to the 90\% confidence interval for a light curve with a power-law
Fourier spectrum $P_{\nu}\propto \nu^{-2}$. The horizontal line is the
average noise level (see text). \label{fourier_power}}
\end{figure}

The average integral flux above 200 GeV observed from PKS\,2155$-$304
is I($>$200 GeV) = 
(1.72$\pm$$0.05_{\rm stat}$$\pm$$0.34_{\rm syst}$)$\,\times\,$10$^{-9}$\,cm$^{-2}$\,s$^{-1}$, 
equivalent to $\sim$7 times the I($>$200 GeV) observed 
from the Crab Nebula
\cite[I$_{\mathrm{Crab}}$;][]{hess_crab}.  Figure~\ref{flux_lc_1min}
shows I($>$200 GeV), binned in one-minute intervals, versus time.  The
fluxes in this light curve range from 0.65 I$_{\mathrm{Crab}}$ to 15.1
I$_{\mathrm{Crab}}$, and their fractional root mean square (rms)
variability amplitude \cite{rms_noise_ref} 
is F$_{\rm var}=0.58\pm0.03$.  This is $\sim$2
times higher than archival X-ray 
variability \cite[Zhang et al.][]{zhang1999,zhang2005}. 
The Fourier power spectrum
calculated from Figure~\ref{flux_lc_1min}
is shown in Figure~\ref{fourier_power}. The error on the power
spectrum is the 90\% confidence interval estimated 
from $10^4$ simulated light curves.  These curves
are generated by adding a random constant to
each individual flux point, where this constant is
taken randomly from a Gaussian distribution
with a dispersion equal to the error of the respective point.
The average power expected when the measurement error
dominates is shown as a dashed line 
\cite[see the Appendix in][]{rms_noise_ref}. There is power
significantly above the measurement noise level up to $1.6 \times
10^{-3}\,{\rm Hz}$ ($600\,{\rm s}$).  The power spectrum also shows that most of the power is at low
frequencies. The grey shaded area shows the 90\% confidence level
obtained by simulating $10^4$ light curves with a power-law Fourier
spectrum $P_\nu\propto \nu^{-2}$ \cite{timmer} and a random Gaussian
error as above. The power spectrum derived from the data is thus
compatible with a light curve generated by a stochastic process with a
power-law Fourier spectrum of index -2. An index of -1 produces too
much power at high frequencies and is rejected. These power spectra are
remarkably similar to those derived in
X-rays \cite[Zhang et al.][]{zhang1999} from the same source.

Rapid variability is clearly visible in substructures that appear in
the light curve, with even shorter rise and decay time scales than found 
in the Fourier analysis.  In
order to quantify those time scales, the light curve is considered as
consisting of a series of bursts, which is common for AGN and
$\gamma$-ray bursts (GRBs).  The ``generalized Gaussian'' shape from
Norris et al.~\cite{norris} is used to characterize these bursts,
where the burst intensity is described by: ${\rm I}(t) = A \exp [
-(|t-t_{\rm max}|/\sigma_{\rm r,d})^\kappa]$, where $t_{\rm max}$ is
the time of the burst's maximum intensity (A); $\sigma_{\rm r}$ and
$\sigma_{\rm d}$ are the rise ($t<t_{\rm max}$) and decay ($t>t_{\rm
max}$) time constants, respectively; and $\kappa$ is a measure of the
burst's sharpness.  The rise and decay times, from half to maximum
amplitude, are $\tau_{r,d}=[\ln 2]^{1/\kappa}\sigma_{r,d}$.  A peak
finding tool, using a Markov chain algorithm \cite{mor02}, selected
five significant bursts. 
A function consisting of a superposition of an identical
number of bursts plus a constant signal
was fit\footnote{The Markov chain burst positions
were used to initialize $t_{\rm max}$ for each burst.  All parameters
are left free in the fit.} to the data.   
The best fit has a $\chi^2$ probability of
20\% and the fit parameters are shown in
Table~\ref{burst_info}. Interestingly, there is a marginal trend for
$\kappa$ to increase with subsequent bursts, 
making them less sharp, as the flare
progresses. The $\kappa$ values 
are close to the bulk of those found by Norris et al.~\cite{norris}, 
but the time scales measured here are two orders of
magnitude larger.

\begin{table}
\caption{The results of the best $\chi^2$ fit of the superposition
of five bursts and a constant to the
data shown in Figure~\ref{flux_lc_1min}. The constant term is 
$0.27\pm0.03 \times 10^{-9}\,{\rm cm}^{-2}\,{\rm s}^{-1}$ (1.1
${\rm I}_{\rm Crab}$).\label{burst_info}}
        \centering
\begin{tabular}{ccccc}
\tableline\tableline
$t_{\rm max}$  & $A$ &$\tau_{\rm r}$ & $\tau_{\rm d}$  & $\kappa$ \\
$[$min$]$ & $[10^{-9}\,{\rm cm}^{-2}\,{\rm
    s}^{-1}]$ & [s] & [s] & \\
\tableline\\
41.0 & 2.7$\pm$0.2 & 173$\pm$28 & 610$\pm$129 & 1.07$\pm$0.20\\
58.8 & 2.1$\pm$0.9  & 116$\pm$53 & 178$\pm$146 & 1.43$\pm$0.83\\
71.3 & 3.1$\pm$0.3 & 404$\pm$219 & 269$\pm$158 & 1.59$\pm$0.42\\
79.5 & 2.0$\pm$0.8 & 178$\pm$55  & 657$\pm$268 & 2.01$\pm$0.87\\
88.3 & 1.5$\pm$0.5 & 67$\pm$44   & 620$\pm$75  & 2.44$\pm$0.41\\
\tableline\\
\end{tabular}
\end{table}

During both the first two bursts there is clear doubling of the flux
within $\tau_{r}$.  Such doubling is sometimes used as a
characteristic time scale of flux variability.  For compatibility with
such estimators, the definition of doubling time, 
$T_2 = |{\rm I}_{ij} \Delta T / \Delta {\rm I}|$, from Zhang
et al.~\cite{zhang1999} was used\footnote{Only values of $T_2$ with 
less than 30\% uncertainty are considered.}.  Here, $\Delta T = T_j - T_i$, 
$\Delta {\rm I} = {\rm I}_j - {\rm I}_i$, $
{\rm {\rm I}}_{ij} = ({\rm I}_j + {\rm I}_i)/2$, with $T$ 
and I being the time and flux, respectively, of any pair
of points in the light curve. 
The fastest $T_2=224\pm60\,{\rm s}$ is
compatible with the fastest significant time scale found by the
Fourier transform. Averaging the five lowest $T_2$
values yields $330\pm40\,{\rm s}$.

The variability time scales of these bursts are among 
\cite[see also][]{MAGIC_501} the fastest ever
seen in a blazar, at any wavelength, and are almost an order of
magnitude smaller than previously observed from this object.  It
should be noted that similar time scales are found with even smaller
binning (e.g. 20\,s) of the H.E.S.S. light curve, and that many checks
of the data quality were undertaken to ensure that the flux variations
cannot be the result of background fluctuations, atmospheric events,
etc.  In addition, all the results have been verified using an
independent calibration method and alternative analysis
techniques. 

\subsection{Spectral Analysis}

\begin{figure}
\plotone{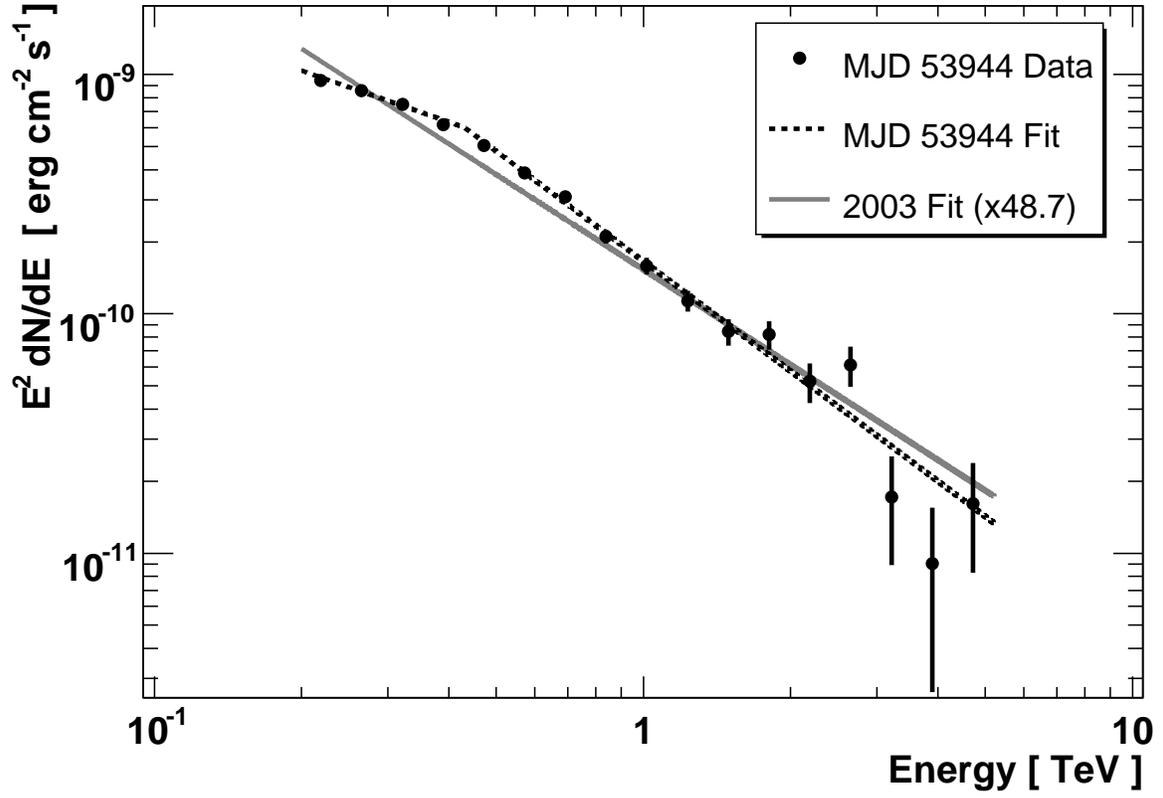}
\caption{The time-averaged spectrum observed
	from PKS\,2155$-$304 on MJD 53944.
	The dashed line is 
	the best $\chi^2$ fit of a broken power law to
	the data.  The solid line represents the fit to
	the time-averaged spectrum of PKS\,2155$-$304 
	from 2003 \cite{HESS_2155A} 
	scaled by 48.7.  Neither spectrum is
	corrected \cite[see, e.g.,][]{HESS_2155B} 
	for the absorption of VHE $\gamma$-rays
	on the Extragalactic Background Light.
\label{avg_spectrum}}
\end{figure}

Figure~\ref{avg_spectrum} shows the time-averaged photon 
spectrum for these data.  The data are well fit, $\chi^2=17.1$ for 13 
degrees of freedom (d.o.f.), by a broken power-law function:

\noindent
$E < E_{\mathrm B}: \frac{\rm dN}{\rm dE} = I_{\circ} \hspace{0.5ex} \left(\frac{E}{\rm 1\,TeV}\right)^{-\Gamma_{1}} \hspace{0.5ex}$ 

\noindent
$E > E_{\mathrm B}: \frac{\rm dN}{\rm dE} = I_{\circ} \hspace{0.5 ex} \left(\frac{E_{\mathrm B}}{\rm 1\,TeV}\right)^{(\Gamma_{2} - \Gamma_{1})} \left(\frac{E}{\rm 1\,TeV}\right)^{-\Gamma_{2}}$,

\noindent
where $I_{\circ}\,$=\,(2.06$\pm$0.16$\pm$0.41) 
$\times$ 10$^{-10}$ cm$^{-2}$\,s$^{-1}$\,TeV$^{-1}$,
$E_{\mathrm{B}}$\,=\,430$\pm$22$\pm$80 GeV,
$\Gamma_{1}$\,=\,2.71$\pm$0.06$\pm$0.10, 
and $\Gamma_{2}$\,=\,3.53$\pm$0.05$\pm$0.10. 
For each parameter, the two uncertainties are the 
statistical and systematic values, respectively.
Fits to the data of either a simple power law 
($\Gamma$\,=\,3.19$\pm$0.02$\pm$0.10,
$\chi^2$\,=\,138, 15 d.o.f) or a power law with an exponential cut-off 
($\chi^2$\,=\,45, 14 d.o.f.) are not acceptable.
The time-averaged spectrum ($\Gamma$\,=\,3.32) 
of \pks\ measured in 2003 \cite{HESS_2155A},
multiplied by the ratio (48.7) of I($>$200~GeV) from the 
respective data sets, is also shown in Figure~\ref{avg_spectrum}.
Despite a factor of $\sim$50 change in flux there 
is qualitatively little difference between the two spectra.
Indeed, fitting a broken power law to the 
current data set, keeping $\Gamma_{1}$ and $\Gamma_{2}$ 
fixed to the values measured in 2003, yields a 
value for $E_{\mathrm{B}}$ consistent with that 
measured in 2003.  The small difference is surprising since a 
change of the spectral shape with varying flux levels, typically
hardening with increased flux, has often
been observed from blazars at X-ray 
energies \cite[see, e.g.,][]{Xray_hardening},
as well as in the VHE domain \cite[see, e.g.,][]{VHE_hardening}.  

The high flux observed from PKS\,2155$-$304 allows the determination of
accurate photon spectra on time scales of the order of minutes.
Therefore, a  simple search for temporal changes of 
the VHE spectral shape within these data was performed.  Spectra were determined 
for consecutive data slices of 28 minutes (1 run), 10 minutes, 
and 5 minutes.  Fitting the time-average spectral shape, allowing only the
normalization ($I_{\circ}$) to vary, to these short-time-scale spectra
yields reasonable $\chi^2$ probabilities.  
Thus, there are no strong indications 
of fast spectral variability.  However,
weak variations ($\Delta \Gamma$\,$<$\,0.2) are not ruled out.
A more sophisticated study of any fast spectral variations 
within these data is beyond the scope of this 
letter and will be published elsewhere.

\section{Discussion}

It is very likely that the electromagnetic emission in blazars 
is generated in jets that are beamed 
and Doppler-boosted toward the observer. 
Superluminal expansions observed with VLBI \cite{piner}
provide evidence for moderate Doppler boosting in PKS\,2155$-$304.
Causality implies that $\gamma$-ray variability on a 
time scale $t_{\rm var}$, with a Doppler factor\footnote{With 
$\delta$ defined in the standard
way as $[\Gamma(1-\beta\cos\theta)]^{-1}$, where $\Gamma$ is the bulk
Lorentz factor of the plasma in the jet, $\beta = v/c$, and $\theta$
is the angle to the line of sight.} ($\delta$), is related to the
radius ($R$) of the emission zone by $R \leq ct_{\rm
var}\delta/(1+z)$. Conservatively using the best-determined rise time 
(i.e. $\tau_r$ with the smallest error) 
from Table~\ref{burst_info} for $t_{\rm var} = 173\pm28\,{\rm s}$ 
(note that this is similar to the
fastest $T_2$) limits the size of the emission region
to $R\delta^{-1} \leq 4.65 \times 10^{12}$ cm $\leq 0.31$ AU.

The jets of blazars are believed to be powered by accretion onto a 
supermassive black hole (SMBH).  Thus accretion/ejection 
properties are usually presumed to scale with the Schwarzschild radius 
$R_{\rm S}$ of the SMBH, where $R_{\rm S} = 2GM/c^2$, which
is the smallest, most-natural size of the system
\cite[see, e.g.,][]{Blandford}.
Expressing the size $R$ of the $\gamma$-ray emitting region 
in terms of $R_{\rm S}$, the variability time
scale limits its mass by $M \leq (c^3 t_{\rm var}\delta/2G(1+z)) R_{\rm S}/R 
\sim 1.6\times10^7 M_\odot \delta R_{\rm S} / R$. 
The reported\footnote{See Wurtz
et al.~\cite{wurtz} and $M_R>-23.1$ (for $h=0.5$) showing the need for
confirmation of this value.}  host galaxy luminosity $M_R=-24.4$
\cite[Table 3 in][]{kotilainen} would imply a SMBH mass of order
1$-$2$\times 10^9M_\odot$ ~\cite{bettoni2003}, and therefore,
$\delta\geq 60-120\,R/R_{\rm S}$.  Emission regions of only a 
few $R_{\rm S}$ would
require values of $\delta$ much greater than those typically
derived for blazars ($\delta$$\sim$10) and come close
to those used for GRBs, which would be a challenge to understand.  For
example, the sub-parsec VHE $\gamma$-ray emitting plasma would have to
decelerate with a high efficiency to accommodate relatively small
Lorentz factors observed at parsec scales \cite{piner}.  It is however
possible that the SMBH mass is over-estimated, reducing the $\delta$
constraint by the same factor, or that the variability has an
origin \cite[e.g., a geometric effect from jet bending 
as discussed in][]{wagner1993} unrelated to the black hole.
Detailed modeling of the spectral energy distribution
of PKS\,2155$-$304, during the multiple VHE 
flares observed by H.E.S.S. in the July 2006 dark period,
including simultaneous multi-frequency data, will
appear elsewhere. 

The VHE variability observed in this particular flaring episode
is the fastest ever observed from a blazar.  While the 
variability is a factor of five times faster than 
previously measured from Mkn 421 \cite{Gaidos_Mkn421}, in terms of the 
light-crossing time of the Schwarzschild
radius, $R_{\rm S}/c$, the variability of PKS\,2155$-$304 is
another factor of $\approx 6-12$ more constraining assuming a
$10^{8.22} M_\odot$ for Mkn 421 \cite{BH_Mass}.
It should also be noted that the 
choice of a $\sim$3 minute variability time scale here is 
conservative and that the light curve is strongly oversampled,
allowing for the first time in the VHE regime a detailed 
statistical analysis of a flare, which shows remarkable similarity to
other longer duration events at X-ray energies.
From such rapid variability one must conclude that 
either very large Doppler factors can
be present in AGN jets, or that the observed variability is not
connected to the central black hole, clearly showing the power
of Cherenkov-telescope arrays in probing the 
internal mechanisms in BL\,Lacs.

\acknowledgments

The support of the Namibian authorities and of the University of Namibia
in facilitating the construction and operation of H.E.S.S. is gratefully
acknowledged, as is the support by the German Ministry for Education and
Research (BMBF), the Max Planck Society, the French Ministry for Research,
the CNRS-IN2P3 and the Astroparticle Interdisciplinary Programme of the
CNRS, the U.K. Particle Physics and Astronomy Research Council (PPARC),
the IPNP of the Charles University, the Polish Ministry of Science and 
Higher Education, the South African Department of
Science and Technology and National Research Foundation, and by the
University of Namibia. We appreciate the excellent work of the technical
support staff in Berlin, Durham, Hamburg, Heidelberg, Palaiseau, Paris,
Saclay, and in Namibia in the construction and operation of the
equipment.

\clearpage

\clearpage

\end{document}